\documentclass[a4paper,10pt,twoside]{cpc-hepnp}

\usepackage{multicol}
\usepackage{graphicx}
\usepackage{booktabs}
\usepackage{amssymb,bm,mathrsfs,bbm,amscd}
\usepackage[tbtags]{amsmath}
\usepackage{lastpage}

\begin{document}

\fancyhead[c]{\small Peking University} \fancyfoot[C]{\small 010201-\thepage}

\footnotetext[0]{Received 15 Junly 2015}

\title{Mechanical analysis of a $\beta=0.09 $  162.5MHz taper HWR cavity\thanks{Supported by  National Basic Research Project (No.2014CB845504) }}

\author{%
      FAN Pei-Liang%
\quad ZHU Feng$^{1)}$\email{zhufeng7726@pku.edu.cn}%
\quad ZHONG Hu-Tian-Xiang\\
\quad QUAN Sheng-Wen
\quad LIU Ke-Xin\\
}
\maketitle

\address{%

 State Key Laboratory of Nuclear Physics and Technology, Peking University, Beijing 100871, China\\
}

\begin{abstract}
One superconducting taper-type half-wave resonator (HWR) with frequency of 162.5MHz, ¦Â of 0.09 has
 been developed at Peking University, which is used to accelerate high current proton ($\sim$ 100mA)
 and $D^{+}$($\sim$ 50mA). The radio frequency (RF) design of the cavity has been accomplished.
 Herein, we present the mechanical analysis of the cavity which is also an important aspect in
 superconducting cavity design. The frequency shift caused by bath helium pressure and Lorenz
 force, and the tuning by deforming the cavity along the beam axis will be analyzed in this paper.
\end{abstract}



\footnotetext[0]{\hspace*{-3mm}\raisebox{0.3ex}{$\scriptstyle\copyright$}2015 Peking University
}%

\begin{multicols}{2}

\section{Introduction}

More and more projects based on high current proton and deuteron linear accelerators are
proposed and emerged to support various fields of science like particle physics, nuclear
physics, and neutron-based physics. Superconducting RF (SRF) technology is mandatory for
high current continuous wave (CW) proton and deuteron linear accelerators. HWR is one of
the best cavity geometries for the acceleration of high-intensity proton and heavy ion
beams in low beta range. Compare to the Quarter-Wave Resonator (QWR), HWRs have no dipole
fields on the beam axis due to the symmetry of EM fields and better mechanical properties.
Recently, various superconducting HWR structures have been developed, of which the taper
type HWR has good mechanical stability and lower maximum surface fields in comparison to
the squeezed type. Ring-shaped center conductor of taper type HWR is proposed by Argonne
National Laboratory \cite{lab01}. Compared to the race-track geometry, it has not only much
lower peak surface magnetic field and significantly higher shunt impedance, but also no
quadrupole electric field component, which is very important for high current beam acceleration.

One superconducting taper type HWR with frequency of 162.5MHz, $ \beta$ of 0.09 has been
developed at Peking University. This cavity is designed to accelerate 100mA proton beam
or 50 mA deuteron beam. The structure of the cavity is shown in Fig. 1. The EM design of
the cavity has been accomplished and the main RF parameters are shown in Table 1.
The diameter of the beam pipe of the cavity is 40 mm. In this paper, we present the
structural analysis of the cavity. The frequency shift caused by bath helium pressure and
Lorenz force, and the tuning by deforming the cavity along the beam axis will be analyzed.

 \begin{center}
\tabcaption{ \label{tab1}  The main RF parameters of the HWR cavity.}
\footnotesize
\begin{tabular*}{80mm}{c@{\extracolsep{\fill}}ccc}
\toprule Parameter & value    \\
\hline
frequency/MHz & 162.5 \\
$\beta$ & 0.09   \\
$B_{peak}/E_{acc}$/(mT/MV/m) & 6.4   \\
$E_{peak}/E_{acc}$ & 5.3   \\
$R/Q_{0}/\Omega$ & 255  \\
$G/ \Omega$& 39.0  \\
\bottomrule
\end{tabular*}
\end{center}

\section{Helium pressure detuning}

When the cavity frequency shifts from the resonance frequency, the cavity stored energy degrades,
and the accelerating field decreases, thus causes the instability of cavity operation. Moreover,
the cavity detuning brings higher requirements for both RF control and RF power level. Microphonics
and Lorentz force detuning are the main factors for cavity detuning. The fluctuations of the liquid
helium bath pressure are the main source of microponics. The pressure sensitivity coefficient, $df/dp$,
is used to characterize the influence of the helium pressure variation on the detuning of the cavity.
Larger $|df/dp|$ may cause serious cavity detuning and instability of cavity operation. The mechanical
design of the SRF cavity needs to have a lower $|df/dp|$.

 \begin{center}
\includegraphics[width=6cm]{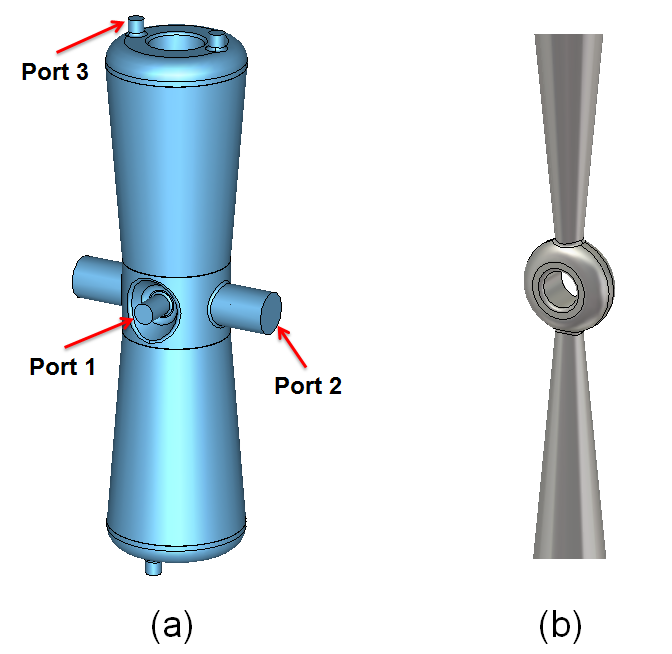}
\figcaption{\label{fig1}  (color online) The structure of the taper type HWR.
(a) The cavity model used in simulation. (b) The ring-shaped center conductor.
 }
\end{center}

 \begin{center}
\includegraphics[width=7.0cm]{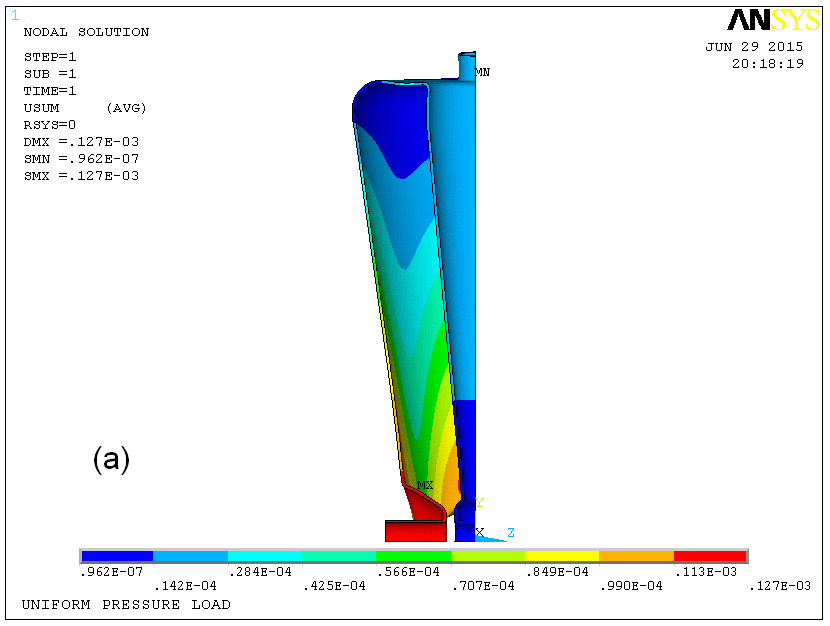}
\includegraphics[width=7.0cm]{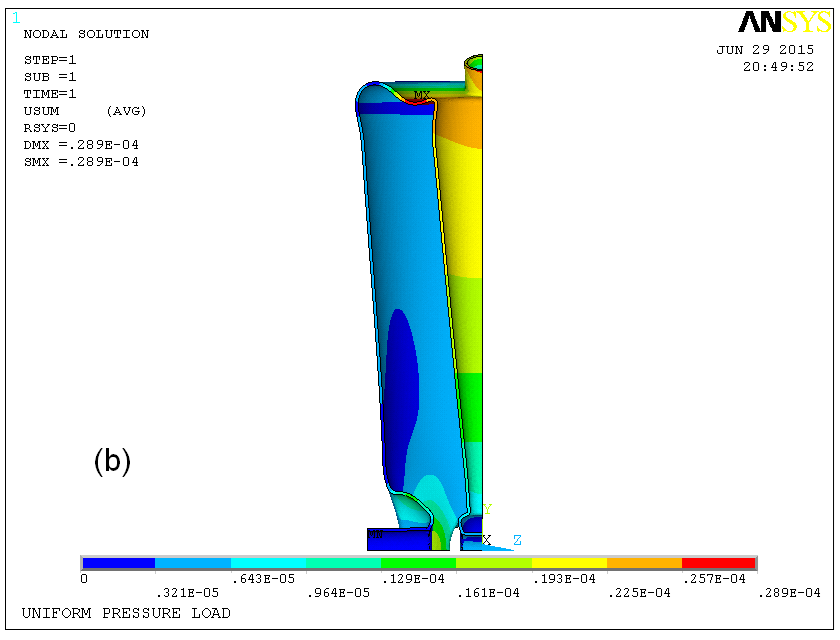}
\figcaption{\label{fig1} (color online) Profile of the deformation result with beam ports free (a) and beam ports fixed (b).
 }
\end{center}

The ANSYS codes \cite{lab02} were used for the simulation. The model was meshed in ANSYS-Workbench and the
simulation was done in ANSYS-APDL. The finite element mesh generation is tetrahedrons, and we used
local mesh refinement scheme in the high field region. The simulation for the helium pressure detuning
was modeled with a series of constant pressure on the surface of the cavity shell.The properties of niobium used in the simulation are list in Table 2.

 \begin{center}
\tabcaption{ \label{tab2}  The properties of niobium used in the simulation.}
\footnotesize
\begin{tabular*}{80mm}{c@{\extracolsep{\fill}}ccc}
\toprule Parameter & value    \\
\hline
Poisson's ratio            & 0.38 \\
Young's modulus/GPa        & 105   \\
Tensile yield strength/MPa & 250  \\
Thickness/mm               & 3.0  \\
\bottomrule
\end{tabular*}
\end{center}

Different boundary conditions at different ports are taken consideration. There are three kinds of ports,
seen in Fig. 1 (a), port 1 is the beam port, port 2 is the coupler port and port 3 is the cleaning port.
Fixed boundary condition and free condition were put on the three kinds of ports. During operation the
boundary condition is close to fixed.

As we known, when a small volume in the high magnetic field region was removed, the equivalent inductance will decrease and
the resonant frequency will increase. In constant, removing a small volume in the high electric field will increase the
equivalent capacitance and reduce the resonant frequency.

The simulation result is shown in Table 3. From the simulation
result we can see that $df/dp$ is -31.5Hz/mbar when all the three kinds of ports are at free boundary conditions. It means
that the deformation in the high electric field plays a major role in the frequency detuning. The deformations under
one atmosphere pressure are shown in Fig. 2. When the beam port is free, the maximum deformation locates in the beam
pipe region, which is in the high electric region. When the beam port is fixed, the maximum deformation locates in the
short end of the cavity, which is in the high magnetic field region, and $df/dp$ is 3.11Hz/mbar in this case.

\begin{center}
\tabcaption{ \label{tab3}  The simulation result for different boundary conditions.}
\footnotesize
\begin{tabular*}{80mm}{c@{\extracolsep{\fill}}ccc}
\toprule Port1 & Port2 &Port3 & df/dp(Hz/mbar)    \\
\hline
free     & free  & free &-31.5\\
fixed    & free  & free &3.11\\
fixed    & fixed & free &3.18\\
\bottomrule
\end{tabular*}
\end{center}

The typical value of $df/dp$ is about 30 Hz/mbar for 1.3GHz elliptical cavity \cite{lab03} and it can be as high as 74Hz/mbar for
medium beta elliptical cavity with the thickness of 5mm \cite{lab04}. The typical $df/dp$ is about -10Hz/mbar for spoke cavity with radial stiffening ribs \cite{lab04}.
The frequency shift caused by changing the helium pressure is about -17.6 Hz/mbar for a squeezed HWR cavity \cite{lab05}. The
pressure sensitivity coefficient is only about 3 Hz/mbar for this taper type $\beta$=0.09 HWR cavity without stiffening
rings. This verifies that taper type HWR cavity has good mechanical properties against pressure detuning.

 \begin{center}
\includegraphics[width=7.0cm]{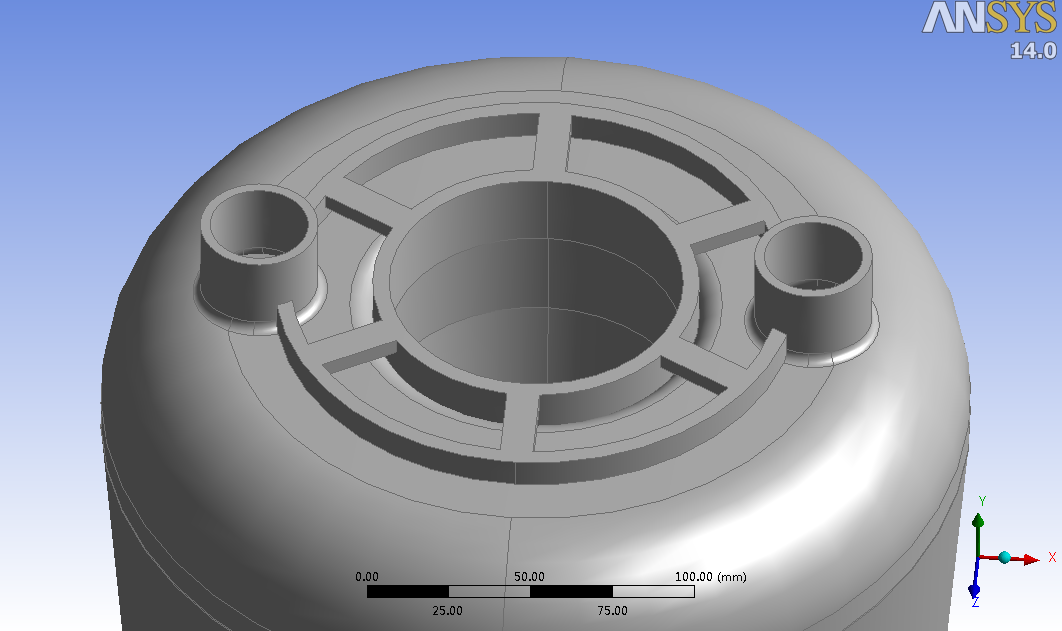}
\figcaption{\label{fig1}  (color online) Stiffening ribs on the short end of HWR cavity.
 }
\end{center}

By adding stiffening ribs on the short end of the cavity, the $df/dp$ can decrease further to 0.01 Hz/mbar when the beam port is fixed.
The structure of the stiffening ribs is shown in Fig. 3. Such low $|df/dp|$ is very good for cavity operation.

\section{Lorenz force detuning}

 Another factor for cavity detuning is Lorentz force. The Lorenz force is the result of the interaction between the rf
 magnetic field in a cavity and the rf wall current.
  The radiation pressure,
\begin{eqnarray}
\label{eq01}
P_{L}\propto \mu_{0} H^{2} - \varepsilon_{0} E^{2}
\end{eqnarray}
causes a small deformation of the cavity shape leading to a change in the resonant frequency:

\begin{eqnarray}
\label{eq02}
\Delta f \propto (\mu_{0} H^{2} - \varepsilon_{0} E^{2}) \Delta V
\end{eqnarray}

Here $\Delta V$ is the change in the volume of the cavity region. Under steady-state condition
this detuning is proportional to the square of the electrical field.

The Lorenz force detuning coefficient $K_{L} $ is used to describe this effect which is defined as

\begin{eqnarray}
\label{eq02}
K_{L}=\Delta f/ E_{acc}
\end{eqnarray}

Here $\Delta f$ is the frequency shift and $E_{acc}$ is the average accelerating electric field.
From Eq. (1) we can see that the direction of the Lorenz force in the electric region and in magnetic
region are opposite which resulting the $K_{L}$ a negative value.

The deformation of the cavity wall with free beam ports is shown in Fig. 4 (a) and
the maximum deformation locates near the high electric field
areas. The relationship between frequency shift and the square
of the electrical field is plotted in Fig. 5. According to the
fitting curve, the Lorenz force detuning coefficient $K_{L}$ is -5.55Hz/(MV/m)$^{2}$  ,
and $K_{L}$=-0.41Hz/(MV/m)$^{2}$ when the beam ports are fixed. Simulation results show
that the taper type HWR cavity also has very low $|K_{L}|$ , which is good for cavity pulse operation.

 \begin{center}
\includegraphics[width=7.0cm]{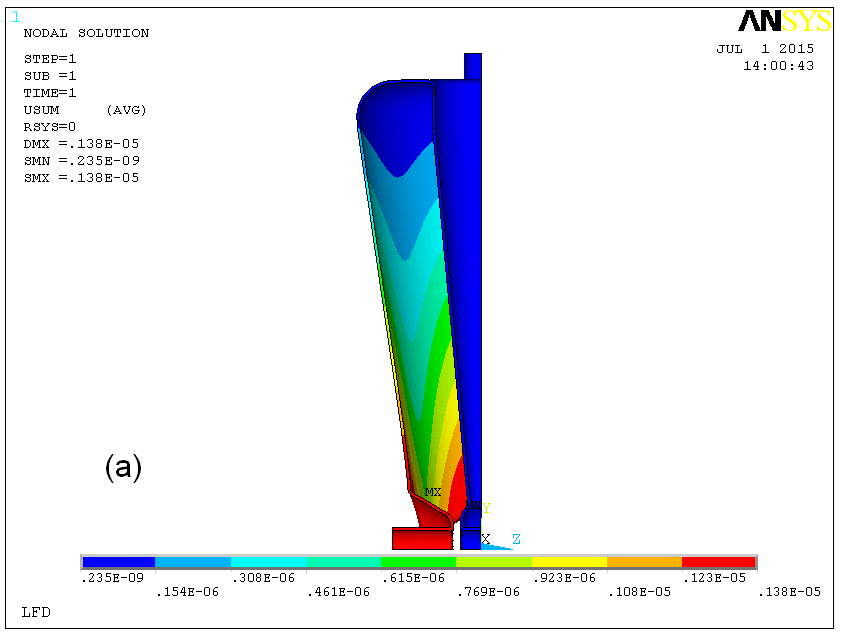}
\includegraphics[width=7.0cm]{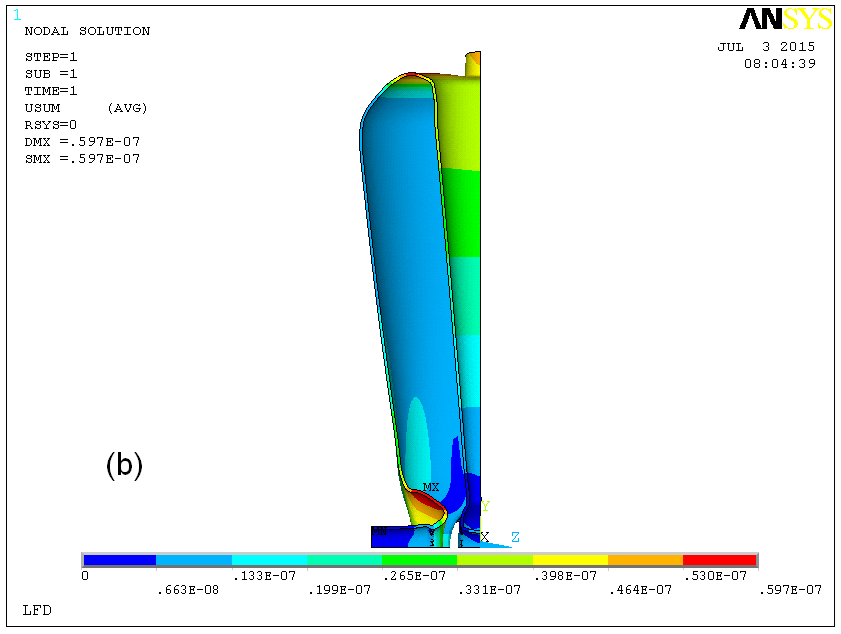}
\figcaption{\label{fig1}  (color online) The deformation of the cavity wall caused by the LFD effect ($E_{acc}$=3.0MV/m)
   The beam port is free (a) and the beam port is fixed (b).
 }
\end{center}

 \begin{center}
\includegraphics[width=7.0cm]{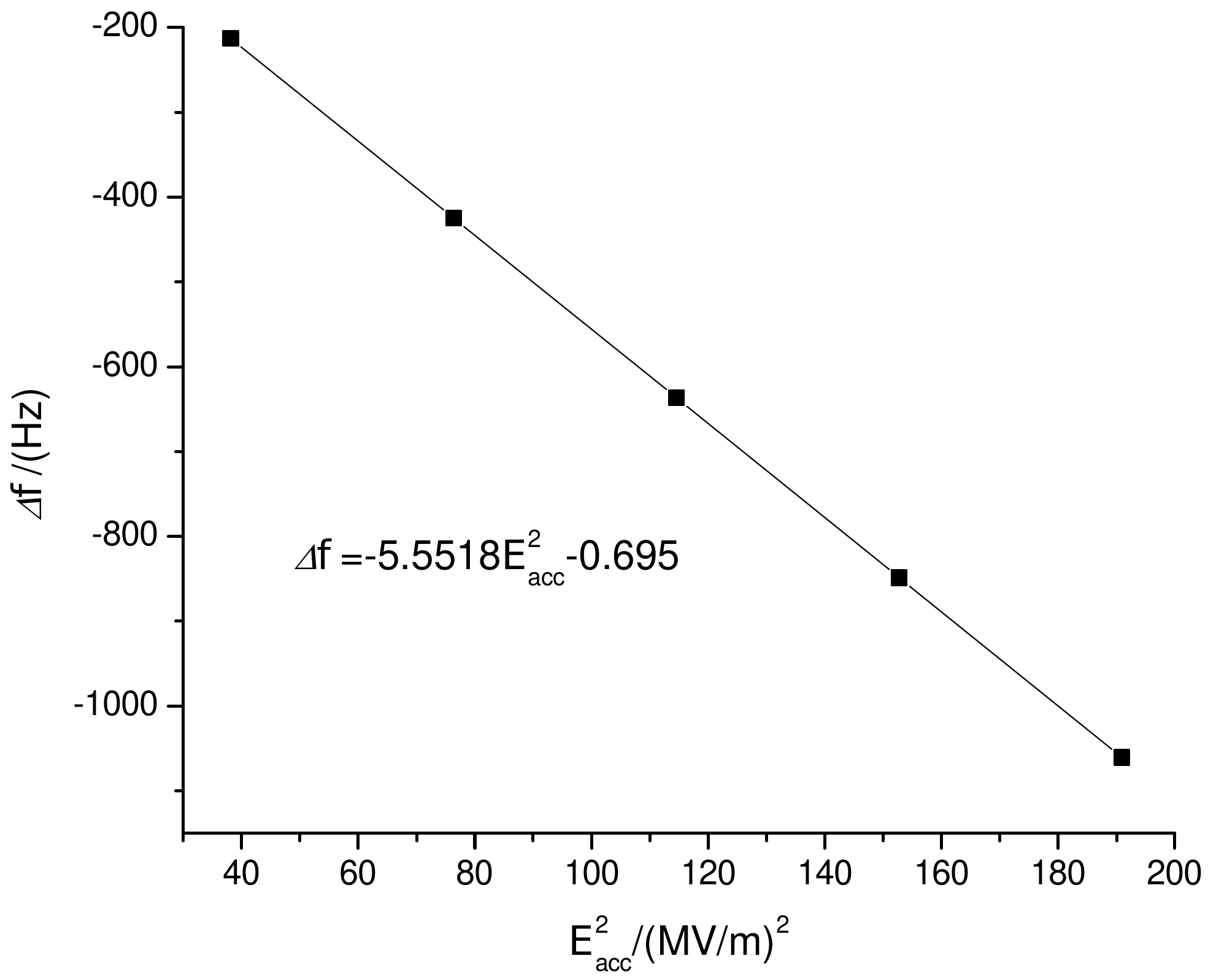}
\figcaption{\label{fig1} The fitting curve between frequency shift and accelerating gradient.
 }
\end{center}

\section{Tuning analysis}

Mechanical tuning method is used to compensate the frequency detuning, and the tuning fore is applied along the beam line, as shown in Fig. 6.

 \begin{center}
\includegraphics[width=7.0cm]{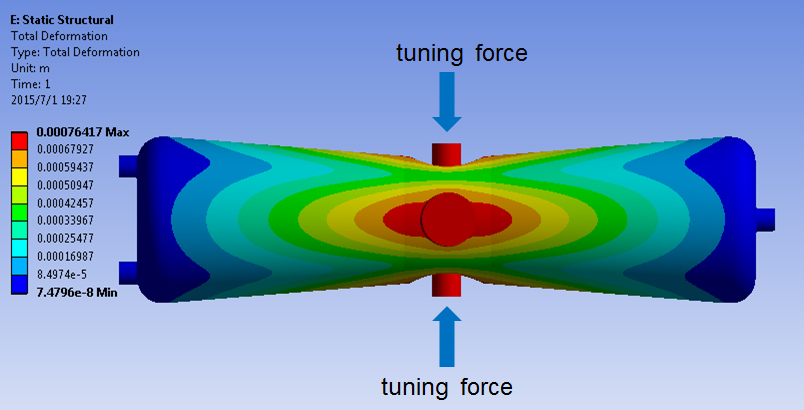}
\figcaption{\label{fig6}  (color online) The deformation of cavity with tuning force.
 }
\end{center}

The simulation result shows that the tuning sensitivity is -253 KHz/mm when the cavity was
squeezed or stretched along the beam line. The displacement is relative to the center of
the cavity. The maximum displacement is about 0.7 mm considering the safe factor larger
than 1.1. In this case, we think the niobium is still in the range of elastic deformation.
In Table 4, we list the displacement and the safe factor. Safe factor 1 is considering
the max equivalent and the safe factor 2 is considering the max shear stress. Therefore,
the tuning range for the HWR cavity is $\pm$ 177KHz.

\begin{center}
\tabcaption{ \label{tab4}  The displacement and the safe factor.}
\footnotesize
\begin{tabular*}{80mm}{c@{\extracolsep{\fill}}ccc}
\toprule Displacement/mm & Safe factor 1 & Safe factor 2    \\
\hline
0.75    & 1.19  & 1.04\\
0.70    & 1.27  & 1.12 \\
0.65    & 1.38  & 1.21 \\
\bottomrule
\end{tabular*}
\end{center}

\section{Summary}

Mechanical analysis is important for the cavity design. Both the helium pressure detuning
and the Lorenz force detuning are studied in this paper. The simulation result shows that
this taper type HWR cavity has better mechanical stability against the helium pressure
fluctuation and Lorentz force detuning than the elliptical cavity, spoke cavity and even
better than the squeezed type HWR cavity. The tuning analysis of the HWR cavity was also
done, which will provide reference for the design of tuner.





\end{multicols}

\vspace{-1mm}
\centerline{\rule{80mm}{0.1pt}}
\vspace{2mm}

\begin{multicols}{2}

\end{multicols}

\clearpage

\end{document}